\documentclass[12pt]{article}

\usepackage{amsmath,amssymb,amsthm,mathrsfs,bbm}
\usepackage{latexsym,amscd,amsbsy,dsfont,amsfonts}
\usepackage{graphicx}
\usepackage[pdftex,colorlinks=true]{hyperref}
\usepackage{caption}

\def\MT#1{{\texttt{#1}}}

\usepackage{bm}
\usepackage{latexsym}
\usepackage[latin1]{inputenc}
\usepackage{amsfonts}
\usepackage{amssymb}
\usepackage{amsmath}
\usepackage{subfigure}

\numberwithin{equation}{section}

\begin{document}

\title{Holographic Lattices and Numerical Techniques}
\author{Tom\'{a}s Andrade \\ \textit{Departament de F\'isica Qu\`antica i Astrof\'isica,} \\ 
\textit{Institut de Ci\`encies del Cosmos (ICCUB), Universitat de Barcelona,} \\ \textit{
Mart\'i i Franqu\`es 1, E-08028 Barcelona, Spain} }

\maketitle

\abstract{In these notes we discuss various methods relevant to the numerical construction of stationary black hole solutions in General Relativity 
with negative cosmological constant. We focus on solutions which explicitly break translational invariance 
along the boundary directions. Within the framework of the gauge/gravity duality, these can be interpreted 
as lattices. They have finite conductivity and thus help us move one step forward towards more realistic scenarios in 
holography.}

\pagebreak

\tableofcontents

\section*{Disclaimer}

These lecture notes will guide us through a short series of lectures on Holography 
and Numerical Relativity, given at the ``School on Numerical Methods in Gravity and 
Holography", held at Universidad de Concepci\'on, Chile, in November
2017. 
%
%
Due to time constraints and mostly my own (misin)formation on the topic, the scope of 
the lectures will by highly pragmatic, in the sense that we will discuss a series of 
methods rather than a rigorous mathematical framework. This means in particular 
that instead of relying on existence theorems, we will let our physical intuition guide us
in the search and construction of a given solution. 
Hopefully, a consistent structure will emerge at the end of the day and we will 
find ourselves with a useful tool-kit which will allow us to handle interesting problems
relevant to cutting-edge research in holography. 

The course will consist of two main parts: a more ``theoretical" one, in which we will
describe the general strategy to tackle a certain class of problems, and a more ``practical"
one, in which we will implement the strategy using one or more algorithms. Although 
the algorithms are (or should be) software-independent, we will perform the implementation 
in Mathematica only. We will briefly discuss the pros and cons of this choice, but the 
students are encouraged to think outside of the Mathematica-box and focus on the 
overall strategies and alternative implementations. \\

I have generated Mathematica notebooks for all the examples, they are available upon request.

\section*{Main references}

The main references for these lectures notes are the books \cite{matlab} and \cite{boyd}. 
We will not follow either of them in any organized way, but when appropriate, we will 
direct the reader to these references to expand on certain topics. We will also point 
out papers about holography which discuss some of the examples treated below.

\section{Introduction}

We do not have a lot of time to talk about the general idea of holography here. 
For the purposes of these lectures, it suffices to say that it is a duality between 
strongly coupled gauge theories and gravitational theories which live on asymptotically 
anti-deSitter (AdS) spacetimes. In practice, the duality allows us to map rather untractable 
QFT problems in the strongly coupled regime to calculationably tractable problems in classical General Relativity
(GR). This fact has fuelled the study of Einstein's gravity in AdS, and more often that not solutions 
which would be discarded as ``too exotic" in the traditional GR sense attract a great deal of attention
thanks to their holographic interpretation. 

This sets the tone of the journey we are about to start: we will construct and probe 
solutions in GR with AdS asymptotics. As it is well-known, Einstein's equations are 
non-linear PDEs, and obtaining analytic solutions is possible only in very limited cases. 
Because of this difficulty and spurred by interesting physical applications, the duality has motivated 
an extensive study of numerical relativity in asymptotically AdS space-times. 
The increasing complexity derived from the incipient ``maturity"  of the field 
makes it unavoidable to resort to numerical techniques. 

One of the applications of the gauge/gravity duality is the study of transport. 
The dual theory picture is familiar: given a certain plasma or material, we turn on a 
perturbation, say, an electric field, and measure the resulting current. The AdS/CFT
correspondence maps this setup to a problem of black holes: 
given a static non-linear solution representing an equilibrium state, 
we study linear fluctuations and compute how the black hole responds.  
However, there is one extra complication. Most black brane solutions, such as 
Reissner-Nordstrom(RN)-AdS, are invariant under translations along the directions parallel to the brane. 
This results in infinite conductivities, which is unrealistic. We know that 
real condensed matter systems contain some form of momentum dissipation, associated 
to the breaking of translational invariance, e.g. the presence of a lattice. 
Therefore, the inclusion of translational symmetry 
breaking solutions emerges in AdS/CFT as a result of the attempt to study transport 
in a more realistic way. 

Studying this problem in holography requires the construction of black holes which 
break translational symmetry {\it explicitly}, i.e. by the inclusion of boundary conditions
which are not translationally invariant along the boundary directions. Needless to say, solutions of this kind 
must be, with very few exceptions, constructed numerically. 
This motivates the main topic of our lectures: constructing and perturbing static black hole solutions
with AdS-like asymptotics which break translational invariance explicitly by the inclusion of space dependent sources. 
Because of the dual interpretation, these are referred to as {\it holographic lattices}. 

It should be stressed at this point that it is also possible to break translations {\it spontaneously}. This type 
of systems have also been extensively studied in the context of holography, and the interplay of the 
spontaneous and explicit structures leads to very interesting effects in black hole physics. We will, however,
restrict ourselves to the explicit case which is technically simpler and yet provides an interesting arena 
in which to explore the joys and dreads of numerical holography. 

These notes are organized as follows: in section \ref{sec:linearODE} we discuss how to solve linear ordinary differential equations
(ODEs) via shooting and relaxation methods.  We then explain how to adapt the previously introduced linear solvers to 
tackle the problem of non-linear ODEs in section \ref{sec:NL_ODE}. In section \ref{sec:linearPDE} we explain how to go about 
solving elliptic, linear, partial differential equations (PDEs). Finally, section \ref{sec:NLPDE} is devoted to the 
solution of non-linear PDEs.

\section{Linear ODEs}
\label{sec:linearODE}

In this section we will discuss the simplest problem of our series: linear,
second order, ordinary differential equation in one variable. This will serve as a building 
block for all the subsequent problems, so a careful study of this setting cannot be overemphasized. 

We will generically denote the linear operator by $L$, the unknown by $f$, and the independent variable by $x$, 
which takes values in the range $x \in [a, b]$. 
In most of the problems of interest, the equations of motion yield singular points 
at the boundaries of our domain (e.g. the conformal boundary of AdS and the black hole horizon). 
We will assume that the singularities are mild enough so that we can make the appropriate redefinitions
to render the unknown function $f$ regular there. 

We will consider two different variants of this problem: boundary value problems (BVP) in which a unique solution is 
obtained after specifying a source term $J$ and boundary conditions for $f$; and eigenvalue problems (EVP), in
which case we will determine the spectrum of the operator $L$ for prescribed boundary conditions.  We thus have, schematically, 
\begin{align}
\label{bvp}
& {\rm BVP}: &\qquad \qquad  &	L f  =  J, &  \qquad  & x \in [a, b] & \\
\label{evp}
& {\rm EVP}: & \qquad \qquad &L f = \lambda f, &  \qquad &x \in [a, b] &
\end{align}
\noindent where $\lambda$ is the eigenvalue. The boundary conditions of interest will generically 
be of the mixed type 
\begin{equation}
	c_1 f'(a) + c_2 f(a) = c_3
\end{equation}
\noindent and similarly for $x = b$. Examples of these 
problems are the computation of conductivities (BVP) and quasi-normal modes (EVP).
Later in this section we will discuss these examples in detail. 
%

\subsection{Shooting vs relaxation}

When confronted with a numerical BVP, there are essentially two main strategies 
we can follow: shooting and relaxation. 

For the purposes of understanding the 
structure of the solutions, i.e. counting its free parameters, shooting is quite 
useful since its easy to visualize. Moreover, integrating the equations using the 
Mathematica function \MT{NDSolve} is straightforward, and a simple algorithm can be 
built using this function. Shooting consists essentially on integrating the equation 
of motion from each of the boundary points towards the interior and, by varying 
a parameter (for example, the value of the derivative of the solution) matching the 
solutions at a point in the interior. 
The drawback of this method is that, for the applications we are interested in, the 
singularities at the end points make it impossible to integrate the equations of motion 
starting exactly at the end points. To avoid this, one must develop a power series solution
to move away from the singularity, and then proceed with the numerical integration from 
$x = a + \epsilon_a$, $x = b - \epsilon_b$.
Typically, there is quite a bit of sensitivity to the choice of $\epsilon_a$ and $\epsilon_b$
(associated among other things to the order to which the power series solution is developed),
and obtaining a reliable solution requires checking the numerical stability under changes of these parameters. 

This makes it quite desirable to work with a method in which the solution can be represented in 
the {\it whole domain}, including the end points. Intuitively, this must be possible since 
at the end of the day the solutions which we are interested in are regular in some appropriate physical sense. Relaxation 
does this for us. At its core, what we do is to discretize the equations \eqref{bvp}, \eqref{evp}, 
by representing $f$ by a set of values on a grid $\{ x_i \}$ which divides $[a,b]$ into small sub-intervals. 
The way in which we split up the interval induces a matrix form for the derivative operator $L$, so we arrive 
at a matrix problem which can be solved by standard techniques. 
Importantly, we {\it do not} evaluate the equations of motion at the boundary of the domain, but instead
replace them by the boundary conditions. This bypasses the singular character of the equations near the 
boundaries an allows us to obtain a solution in the entire domain. 
The name 'relaxation' will become clearer in the context of non-linear problems, but we will 
stick to this name for now.

\subsection{Shooting}
\label{sec:shooting}

Let us now discuss in detail how the shooting strategy works for the type of problems we described above. For applications
in GR, a key point is that the equations of motion are singular at the end points, so a power series solution is 
required in order to move away from the end points. In practice, this means that the domain of integration of the numerical 
problem is reduced from $[a, b]$ to $[a + \epsilon_a , b - \epsilon_b]$. Let us denote the power series solution of the asymptotic equations 
of motion by $f^S_a(x)$ and $f^S_b(x)$. We assume that these are polynomials, with possibly logarithmic branches, of the form
\begin{equation}
	f^S_a(x) = \sum_{i} c_i (x-a)^i + \ldots 
\end{equation}
\noindent and similarly for $x = b$. The ellipses denote possible logarithmic terms, which can appear in the expansion 
depending on the indices of the Frobenius problem. 

Here comes the key ingredient of shooting: while most of coefficients 
$c_i$ are determined by the boundary conditions and the asymptotic equations, for well-posed problems there should be  
enough freedom so that we can define multi-parametric families of solutions by integrating the equations of motion from each end
of the interval. Let us call the set of free parameters near the ends point $x = a, b$ by  $\{ A \} $ and $\{ B \} $, respectively.
Upon integration from both ends towards at matching point, which we denote by $x_M$, we can construct 
numerical approximations of the solutions to the respective initial value problems, which we call $\bar f_a (x)$ and 
$\bar f_b (x)$. For the second order equations we are interested in, the matching conditions have the form 
\begin{equation}\label{matching ODE}
	\bar f_a (x_M) = \bar f_b (x_M), \qquad \frac{d}{d x}\bar f_a (x_M) = \frac{d}{d x} \bar f_b (x_M)
\end{equation}
We can think of these equations as providing relations among the initial data sets $\{ A \} $ and $\{ B \} $ such 
that the equations of motion and boundary conditions are simultaneously satisfied. 

It is then easy to count the number of free parameters of the resulting solutions. Denoting the total number of 
free parameters by $N_p$, we distinguish three possibilities: 

\begin{itemize}

\item $N_p < 2$, the equations of motion do not have solution.

\item $N_p = 2$, the equations of motion admit a discrete series of solutions. 

\item $N_p >2$, the equations of motion admit a family of solutions with $N_p - 2$ parameters.

\end{itemize}

It is worth emphasizing that both BVP and EVP can be handled with shooting. In order to deal with the later, we 
simply include the eigenvalue among the parameters for which we shoot for, and obtaining a solution of the matching 
equations immediately provides an eigenvalue. 

It should be noted that an additional numerical algorithm is required to solve the matching conditions \eqref{matching ODE}. A popular 
one being Newton-Raphson, usually implemented by the compiled Mathematica function \MT{FindRoot}. This algorithm requires the specification 
of a ``seed" or starting point to initialize the algorithm. The converge of this method is a priori not guaranteed and strongly depends
on the proper selection of the seed. It is strongly advisable to obtain seeds by proximity to known (perturbative or the like) solution. 
Avoid as much as possible to ``shoot in the dark". 

It is worth noting that both \MT{NDSolve} and \MT{FindRoot} can be used to solve multiple coupled equations, so the extension 
of the above algorithm to the case of more than one equation is straightforward. It should be stressed however that in this 
case the number of free parameters over which to shoot increases rapidly, and with it the importance of initializing the 
procedure with a good seed.

\subsection{Relaxation}
\label{sec:relaxation}

\begin{figure}[h]
\centering
\includegraphics[scale=0.5]{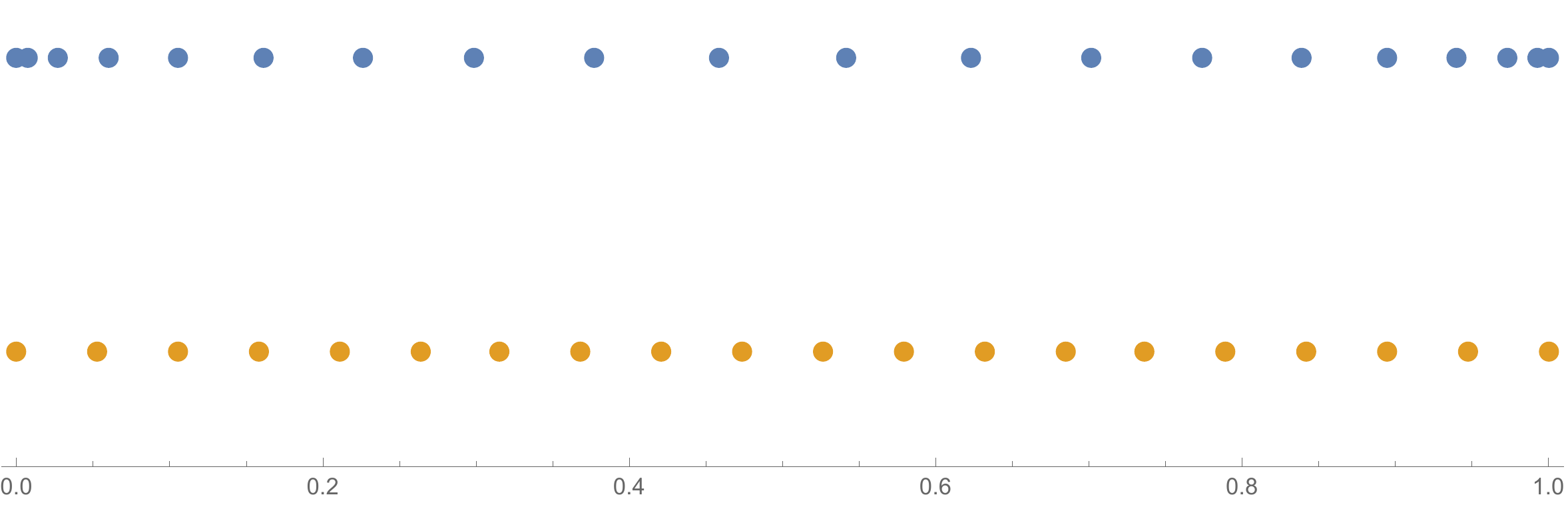}
\caption{\label{fig:ChebHom} Chebyshev (blue) and equispaced grid (yellow) of 20 points.}
\end{figure}

We will now discuss the ``relaxation" procedure. By this we abuse the terminology and
mean in practice that we use a discretization method for the interval $[a,b]$ which begins by
introducing a grid, i.e., a set of points 
\begin{equation}
	\{ x_i \}, \qquad i = 1..N+1 , \qquad {\rm such \, that \,} \quad  x_1 = a, \quad x_{N+1} = b
\end{equation}
For a given grid, it is possible to assign to a derivative operator $\partial_x^n$ a corresponding $(N+1) \times (N+1)$ matrix which 
computes the discretized $n$-th derivative of order $k$, $D_n^{(k)}$. You can look this up in \cite{matlab, boyd}, 
but fortunately Mathematica has them built-in for the most popular grid choices. Most notably, 
homogeneous and Chebyshev grids, with the options of using periodic boundary conditions 
are easily obtained. Moreover, for a Chebyshev grid, one can choose as the order of the derivative to be pseudospectral, 
which gives the most dense derivative operator. We depict a Chebyshev and homogeneous grid in Fig. \ref{fig:ChebHom}.
You can think of this as an operator of infinite order, in the sense that for infinitely differentiable functions, the 
pseudospectral approximation gives exponential convergence with $N$. Consult \cite{matlab, boyd} for 
a thorough discussion on spectral methods. 

Consider an operator $L$ which contains the term $\sim \ell_2(x) \partial_x^2 f$. This operator is to be 
replaced by the discrete version $M_2 \cdot D_2^{(k)}$, where $M_2$ is the diagonal matrix whose entries 
are  $\ell_2(x_i) $, $D_2^{(k)}$ the second derivative matrix of order $k$ and $\cdot$ the usual matrix multiplication. 
Following this procedure,  it is clear how to write the equations of motion in discrete form. Schematically, we write
$M_L \cdot \vec f = \vec J$, where $M_L$ is the matrix which represents the differential operator $L$, and 
$\vec f$ and $\vec J$ are vectors which contain the unknown function and source evaluated at the grid points.  
Now we need to impose the boundary conditions. To this end, we simply replace the first and last rows of the matrix $M_L$, and
the first and last entries of $\vec J$ with the corresponding boundary conditions. You can convince yourself that this 
is precisely what we do in order to enforce the boundary conditions in the continuous problem. 

The resulting matrix problem can be solved by standard methods. In Mathematica, the command we need is \MT{LinearSolve}. 
Once a solution is obtained, we always need to check for the robustness of the answer obtained under the change 
of the grid size. Any results that do not survive this test should be discarded as spurious. 

While this method requires more slightly more knowledge than shooting, this extra work usually pays off since we are able to 
obtain an approximation to the solutions which includes the end points, and as such, is independent of the various cut-offs 
that appear in shooting algorithms with singular points. This is particularly useful when constructing solutions that need to be 
later perturbed, which involves employing the solution itself as the building blocks of the differential operator of the perturbed problem. 

A slight modification of the method described above is required to solve EVPs. Namely, we have to identify the matrices which participate
in the discrete {\it generalized EVP}, which has the form
\begin{equation}\label{linear EVP}
	(M_0 +  \lambda M_1) \vec f = 0 
\end{equation}
The matrices $N_2$ and $M_2$ are (possibly degenerate) matrices corresponding to the discrete versions of 
the associated continuum EVP. Generalized EVP can be handled in Mathematica with 
\MT{Eigenvalues} or \MT{Eigensystem} -- the former only computes the eigenvalues with the latter also provides the eigenfunctions. 

In some cases, it is not possible to write the EVP in this form, and what we 
obtain instead is a {\it quadratic EVP} of the form 
\begin{equation}\label{quad EVP}
  	(M_0  +  \lambda M_1  + \lambda^2 M_2) \vec f  = 0 
  \end{equation}  
Paying the price of doubling the size of the relevant matrices, we can reduce the quadratic problem 
\eqref{quad EVP} to the form \eqref{linear EVP}, by solving $L(\lambda) z = 0$, with 
\begin{equation}
 L(\lambda) = 
   \lambda	\left( \begin{array}{cc}
	M_2 & 0 \\ 
	0 & 1
	\end{array}  \right)  + 
	\left( \begin{array}{cc}
	M_1 & M_0 \\ 
	-1 & 0
	\end{array}  \right) 
\end{equation}
\noindent with eigenvector $z = (\lambda \vec f, \vec f)$.

The extension to systems of equations is straightfoward. A general linear system has the schematic form 
\begin{align}
	L_{ij} \phi_ j &= J_i
\end{align}
\noindent where $L_{ij}$ are differential operators, $J_i$ sources and $\phi_i$ the unknowns. It is clear that we 
can write this in matrix form separately of each equations and repack the full system as a large matrix constructed from the blocks
of each individual equation with unknown $\Phi = \vec \phi$. To this end, the relevant Mathematica command is 
\MT{ArrayFlatten}, which allows us to assemble matrices in blocks.

\subsection{Example 1: Conductivity of RN}

Let us consider a concrete example to fix ideas. In order to keep math simple, we will solve 
a single equation, which appears in the context of perturbations of the translationally invariant 
AdS-RN solution. We will solve this problem by shooting using \MT{NDSolve}. 

The perturbations which control the conductivity of the RN solution at chemical potential $\mu$  can 
be written as
\begin{equation}
	f \delta A_x'' + f' \delta A_x' + \left( \frac{\omega^2}{f} -  (A_t')^2 \right) \delta A_x = 0
\end{equation}
Here, $\delta A_x$ is the perturbation of the $x$ component of the gauge field with frequency $\omega$, and 
\begin{align}
	f &= (r-r_H) \left( r + r_H + \frac{r_H^2}{r} - \frac{r_H \mu^2}{4 r^2} \right) \\
	A_t & = \mu \left( 1 - \frac{r_H}{r} \right)
\end{align}
This equation has singular points at the horizon, $r = r_H$, the conformal boundary $r= \infty$ and singularity $r = 0$.
To perform the numerics it is convenient to introduce the coordinate $z = r_H/r$ so that the conformal boundary is at 
$z= 0$ and the horizon at $z =1$.
The near boundary expansion of the gauge field is given by 
\begin{equation}
	\delta A_x = \delta A_x^{(0)} + z \delta A_x^{(1)} + O(z^2)
\end{equation}
These accounts for two free parameters in the UV.
As usual, near the horizon, the field behaves as 
\begin{equation} \label{NH exp Ax}
	\delta A_x  \approx (1-z^2)^{\pm i \omega /(4 \pi T)} ({\rm regular \, terms})
\end{equation}
Imposing ingoing boundary conditions means that we need to keep the branch with the negative sign in \eqref{NH exp Ax}. 
In order to do this, it is convenient to redefine $ \delta A_x $ in such a way that the new field admits a regular power series, for example
\begin{equation}
	\delta A_x  =:  (1-z^2)^{- i \omega /(4 \pi T)} \hat{ \delta A_x  }
\end{equation}
With this replacement, we obtain a linear second order ODE for $\hat{ \delta A_x  }$, which involves linear and quadratic 
powers of $\omega$.
By construction we have 
\begin{equation}
	\hat{ \delta A_x  } = \hat{\delta A}_x^H + O(1- z)
\end{equation}
This then yields one free parameter in the IR. 
Using the AdS/CFT dictionary, we can write the conductivity as
\begin{equation}
	\sigma(\omega) = \frac{\delta A_x^{(1)}}{i \omega \delta A_x^{(0)}}
\end{equation}
After imposing boundary conditions at the horizon, this becomes a function of $\omega$ only. To see this, 
recall that the equation of motion is linear, so we are free to scale away one of the free parameters, e.g. 
$\delta A_x^{(0)} =1$. For a given $\omega$, the two matching conditions fix the two remaining free parameters 
$\delta A_x^{(1)}$ and $\hat{\delta A}_x^H$, which determines the solution uniquely. \\

\begin{figure}[h]
\centering
\includegraphics[scale=0.5]{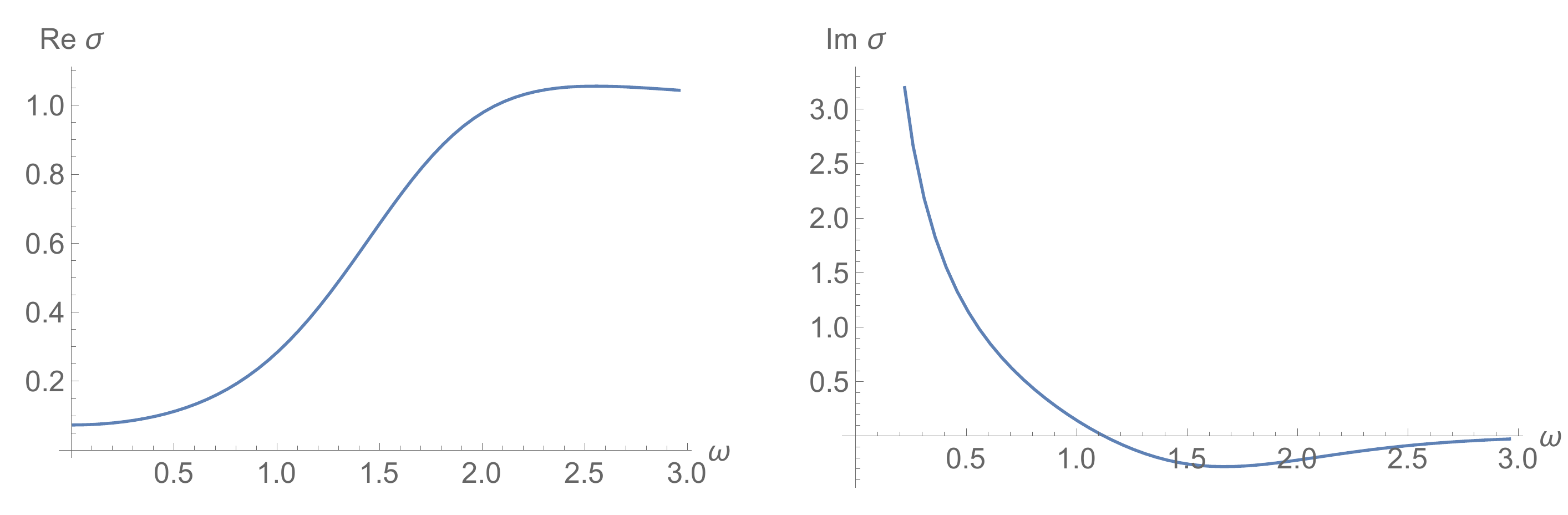}
\caption{\label{fig:sigmaRN} Real and imaginary parts of the conductivity in RN for $T/\mu = 0.2$}
\end{figure}

Solving the numerical problem, we obtain the conductivity plot in Fig \ref{fig:sigmaRN}. Note the pole in the 
imaginary part of $\sigma(\omega)$ near $\omega = 0$. This is the result of having a translational invariant system.  \\

{\bf Exercise}: in this simple setting, we can compute the conductivity using simply an initial value problem. Understand how and compare 
your results with the ones obtained with the method above.

\subsection{Example 2: QNM of AdS-Schwarzschild in 5D}

Let us now discuss an example involving an EVP. The wave equation that governs the 
zero momentum gravitational modes of a five dimensional, planar AdS black hole takes the form 
\begin{equation}
	h_{xy}'' + \left( \frac{f'}{f} - \frac{1}{r} \right) h_{xy}' + \left( \frac{\omega^2}{f^2} - \frac{2 f'}{r f} \right) h_{xy} = 0 
\end{equation}
Here, $h_{xy}$ represents the tensor perturbations of the metric with frequency $\omega$. The emblackening factor reads
\begin{equation}
	f = r^2 \left( 1 - \frac{r_H^4}{r^4} \right)
\end{equation}
Once again we introduce the radial coordinate $z = r_H/r$. The near boundary asymptotics are of the form 
\begin{equation}
	h_{xy} = h_{xy}^{(0)} + \ldots + z^4 h_{xy}^{(4)} 
\end{equation}
The boundary conditions which give the QNM associated to the $\langle T_{xy} T_{xy} \rangle$ correlator are 
$h_{xy}^{(0)} = 0$. In addition, we need to impose ingoing boundary conditions, which 
are identical to the ones described in the previous example. Performing the replacement
\begin{equation}\label{h to ig}
  h_{xy}  =:  (1-z^2)^{- i \omega /(4 \pi T)} \hat h
\end{equation}
\noindent we obtain a second order ODE which involves up to quadratic powers of the frequency. Note that 
in order to impose regularity at the horizon, we can impose the mixed boundary condition which results from 
expanding the wave equation at $z = 1$.
Upon discretization, we obtain a matrix EVP of the form \eqref{quad EVP}, so we can solve this 
by the method described above. Note that the appearance of linear and quadratics powers of $\omega$ 
is due to the redefinition \eqref{h to ig}, and can usually be avoided by using from the onset 
Eddington-Finkelstein coordinates. 

\begin{figure}[h]
\centering
\includegraphics[scale=0.5]{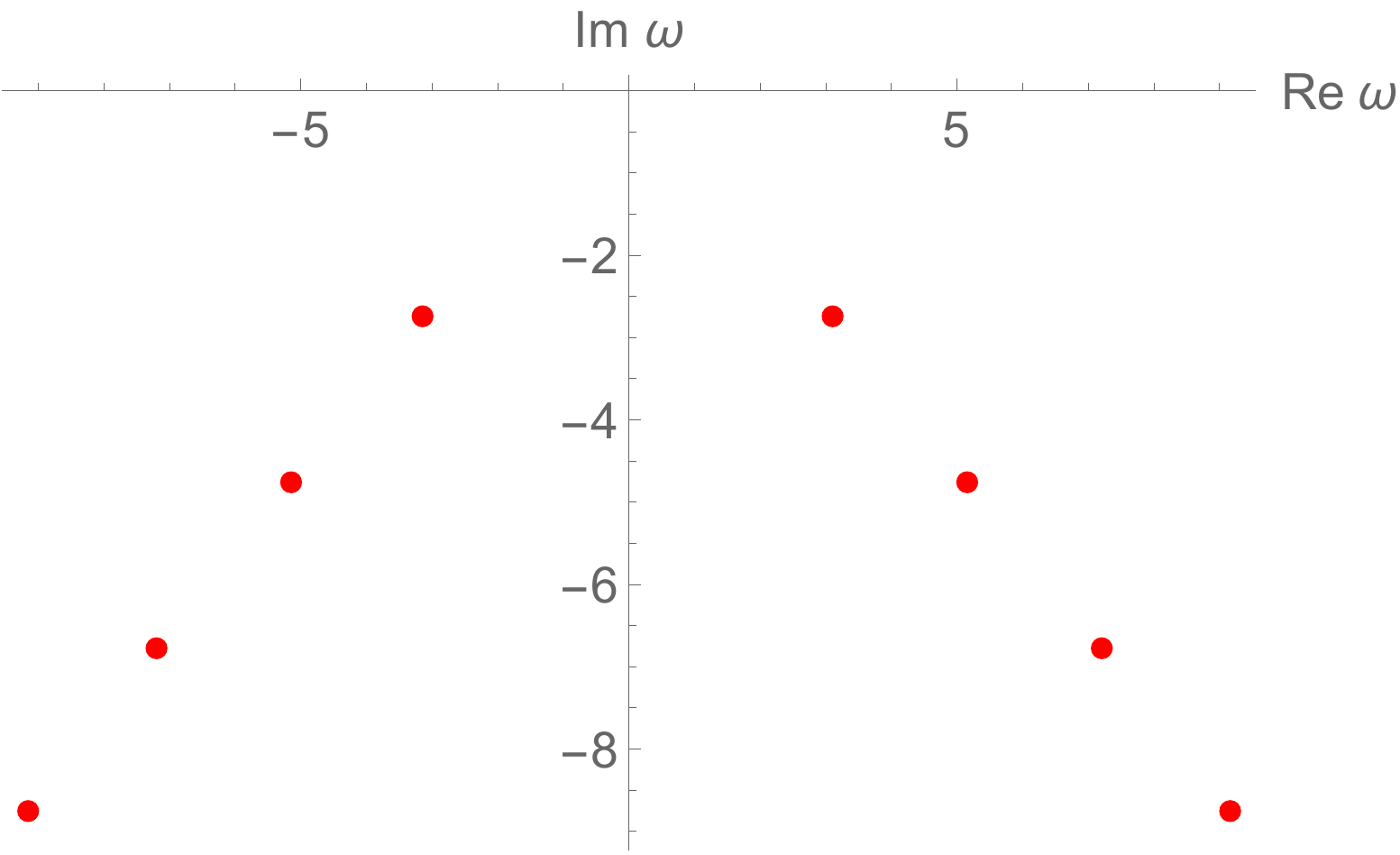}
\caption{\label{fig:QNM AdS5} QNM in the complex plane for an AdS Schwarzschild black brane in 5D. This plot was generated with $n=20$ and pseudospectal methods.}
\end{figure}

We discretize using spectral methods (note that the target functions do not include logs 
since these are proportional to $h_{xy}^{(0)} $ which we are setting to zero). We obtain for the first QNM 
\begin{equation}
	\omega = \pm 3.1194 - i 2.7466, \quad  \pm 5.1695 - i 4.7635, \quad 
	\pm 7.1879 - i 6.7696
\end{equation}
\noindent which is in good agreement with the results of \cite{Starinets:2002br}. We depict these results in the complex plane in Fig \ref{fig:QNM AdS5}. 

\section{Non-Linear ODEs}
\label{sec:NL_ODE}

We now move on to the topic of non-linear ODEs. From the 
pure maths point of view, these type of problems are in fact significantly harder than 
their linear counterparts, in the sense that very little is known about existence-uniqueness 
theorems or similar formal results. Indeed, the analytical techniques to handle non-linear
ODEs are very limited, and usually one can only attempt to solve such systems perturbatively. 
However, if we have enough intuition that a certain solution might exist, then 
solving non-linear ODEs numerically is only marginally harder than solving linear ones. 

If one wishes to attempt shooting, then for most systems direct integration with plain \MT{NDSolve} 
can handle linear and non-linear ODEs equally well, so the shooting method 
of Sec. \ref{sec:shooting} essentially follows through without substantial modification. Once again, 
the hardest part of the algorithm is to find a good initial seed, and usually one resorts to some 
perturbative scheme to produce them.

\subsection{Newton-Raphson}

On the other hand, the true power of the matrix problems begins to emerge when considering non-linear problems. 
This is because, thanks to the iterative NR method, solving a non-linear problem essentially amounts to solving 
a linear problem many times. Consider a non-linear equation for the unknown $f$,
\begin{equation}\label{NL eq}
	E[f] = 0
\end{equation}
Here we use a notation that makes it explicit that we are thinking about this as a differential operator $E$
acting on the function $f$.
Then, the NR method consists of obtaining a series of configurations $f_i$ given an initial seed $f_0$ with updates
\begin{align}
\label{step f}
    f_{i + 1} = f_i + \delta f_i  \\
\label{lin NR}
    E[f_i] + \delta E _{f_i} [\delta f_i]   = 0    
\end{align}
Here, $\delta E _{f_i} [\delta f_i] $ is the linear operator which results from linearising $E$ around the 
background $f_i$ via $E[f_i + \delta f_i] = E[f_i] + \delta E _{f_i} [\delta f_i]  + O(\delta f^2)$, so it can be naturally 
thought as a linear operator acting on the perturbation $\delta f_i$. We declare victory when a configuration $f_{V}$ 
satisfies both $f_{V+1} - f_{V}$ and $E[f_V]$ are ``small" in some appropriate sense

The reader is probably anticipating how to proceed: upon discretization, \eqref{lin NR} turns into a 
BVP which requires us to invert the matrix representing $\delta E$ in the discrete space. Physically, 
the invertibility of the operator stems from the correct choice of the boundary conditions, so these must be enforced carefully. 
A simple way to do it is to write the full system -- including the boundary conditions -- and then linearize. Moreover, 
the proper choice of a seed to initialize the iterative procedure can be the difference between success and failure. 

In sufficiently complicated scenarios, the inversion of the matrix $\delta E$ is computationally costly, 
either because it takes too long or it overflows the memory of our machine. In such cases, it is recommended to use 
some form of a pseudo-Newton method. This refers to iterative methods of this 
kind in which some approximation is used to avoid computing the fully fledged form of $\delta E _{f_i}$, 
in addition to a modification of the step, so that \eqref{step f} becomes $ f_{i + 1} = f_i + \alpha \delta f_i $
with $\alpha <1$.
Intuitively, we do not need to approach the solution by the ``steepest" possible descent, but use 
``some" descent. 
Popular examples are the Broyden method or other types of {\it preconditioning}. 
In the former, by updating $f_i$ in addition to $\delta E _{f_i}$,  we avoid the computation of the inverse. The optimal 
choice for the update was studied by Broyden and you can find it in \cite{boyd}.
In the latter, we use different approximation schemes to evaluate
$E[f_i]$ and $\delta E$, for example, pseudospectral and nearest neighbours respectively. This is accompained by 
the choice of a particular value of $\alpha$.
In our examples we will stick to strict NR. You can find more information about  alternative pseudo-NR methods
in \cite{boyd}. 

It is clear from the above discussion that we can handle systems of non-linear coupled equations by linearizing 
first and then following the block-form reasoning discussed in section \ref{sec:relaxation}.

\subsection{Example 3: $u_{xx} = e^u$}

Our first example of a non-linear ODE is the problem 
\begin{equation}
\label{NL BVP 1}
	u_{xx} = e^u , \qquad - 1 < x < 1 , \qquad u(\pm 1) = 0
\end{equation}
which was taken from \cite{matlab}. This is sufficiently simple that we can 
write down the linear problem explicitly using very little space. We obtain the following BVP 
\begin{equation}
(\partial_x^2  - e^{f_0} ) \delta f = e^{f_0} - \partial_x^2 f_0
\end{equation}
The boundary conditions for $\delta f$ clearly depend on our choice of seed. For example, if the seed 
satisfies the bc, then the subsequent $\delta f$'s should have $\delta f (\pm 1) = 0$. 
It is however advisable to write down a code which can linearize a general problem, we will discuss this in the 
afternoon session. 

\begin{figure}[h]
\centering
\includegraphics[scale=0.5]{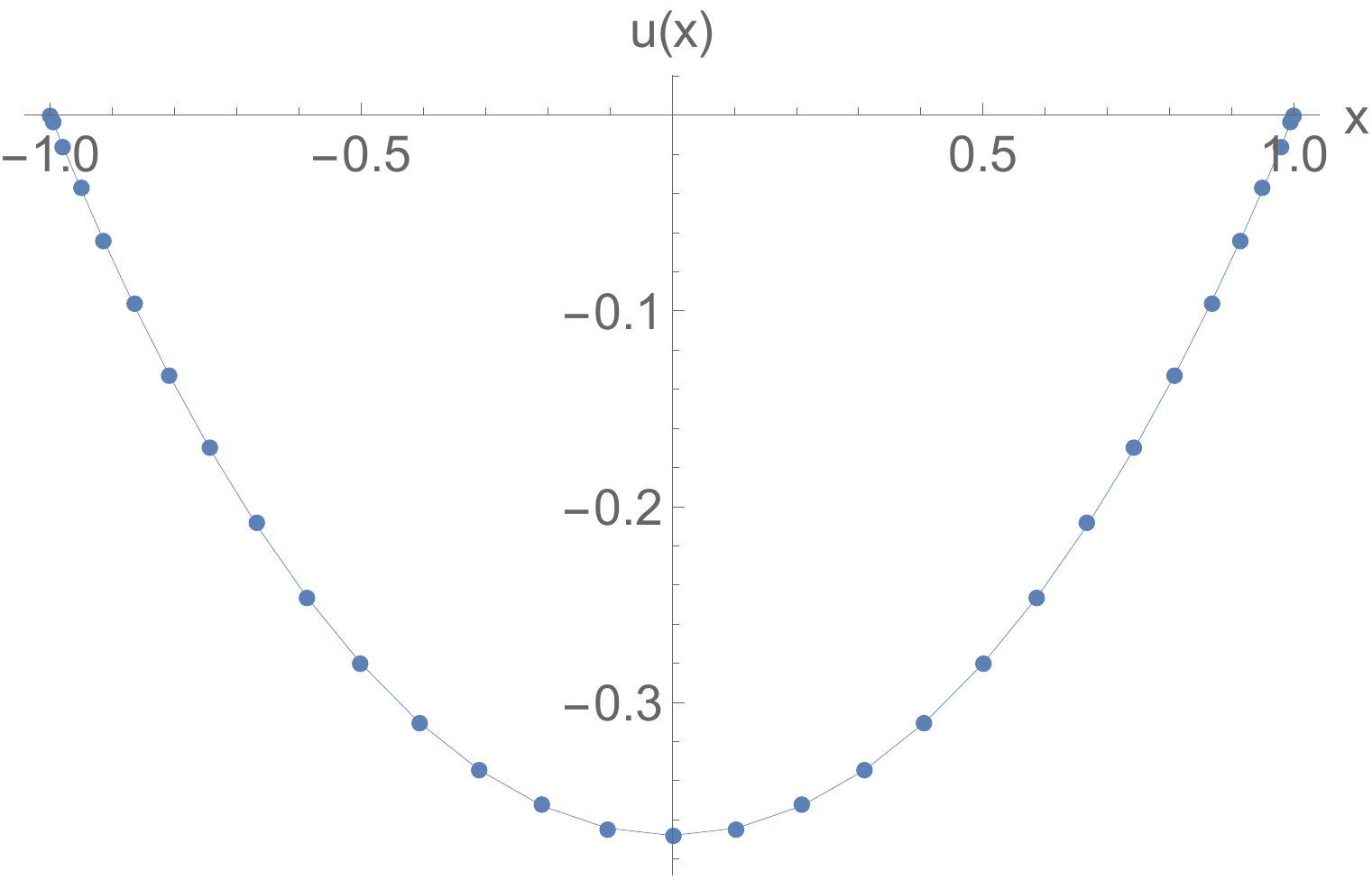}
\caption{\label{fig:SNL_ODE} Solution to the BVP \eqref{NL BVP 1}. We find this solution with $n= 31$ grid points on a Chebyshev 
grid, after 4 iterations. We get $u(0)= -0.368056$ in perfect agreement with \cite{matlab}.}
\end{figure}

We plot our numerical solution in Fig \ref{fig:SNL_ODE}. 

\subsection{Example 4: coupled equations for Q-lattices}

We will now finally deal with a problem which appears in the context of breaking of translational 
invariance in AdS/CFT. This takes place in the context of AdS gravity coupled to complex yet neutral scalar field $\psi$ and 
was introduced in \cite{Donos:2013eha}. We consider an action functional of the form
\begin{equation}
	I = \int d^4 x \sqrt{- g} \left[ R + 6 - \frac{1}{4} F^2 - |\partial \psi |^2 - m^2 |\psi|^2 \right]
\end{equation}
Note that, despite being complex, the scalar $\psi$ is not charged under the $U(1)$. This is only 
a way to introduce a global $U(1)$ which we will allow us to simplify the equations of motion. To 
simplify the UV asymptotics we set 
\begin{equation}
	m^2 = - 2
\end{equation}
The ansatz for the metric and scalar are of the form 
\begin{align}
	ds^2 &= \frac{1}{z^2}\left[ - (1- z) U(z) dt^2 + \frac{dz^2}{(1- z)U(z)} + V_1(z) dx^2 + V_2(z) dy^2 \right] , \\
  \psi &= e^{i k x} z \chi (z) , \\
	A &= (1- z) a(z) dt
\end{align}
Thanks to the global $U(1)$ symmetry associated to the phase change of $\psi$, we can reduce the equations of motion to 
ODEs despite the explicit $x$-dependence  (hence the name Q-lattice). The equations of motion 
are too long to transcribe here, but it suffices to say that we can reduce them to 4 non-linear 2nd order coupled ODEs for
$V_1$, $V_2$, $a$ and $\chi$ in addition to a first order constraint for $U$.

The problem at hand differs from the previous examples in that in this case we need to solve a system of coupled 
(non-linear) equations. While this does not present more trouble than bookkeeping, it is important to 
familiarize with these coupled systems of equations since they are generically what we will have to deal with
when constructing numerical black holes. As mentioned, the way to handle this is to first linearize the full problem and
then cast is in block form.  

\begin{figure}[h]
\centering
\includegraphics[scale=0.5]{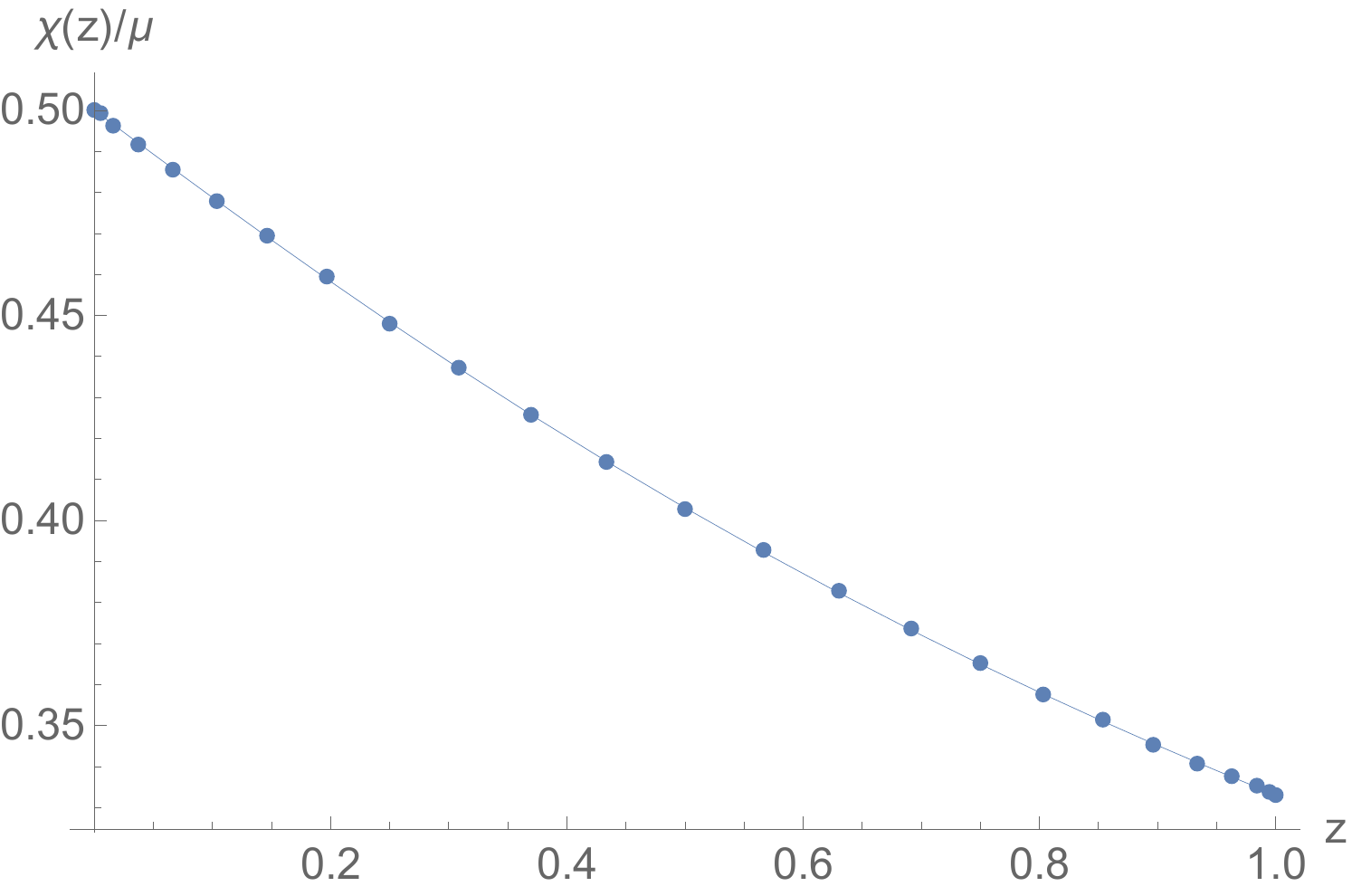}
\caption{\label{fig:chi_QL} Profile of $\chi$ for a Q-lattice with $T/\mu = 1$, $k/\mu = 1$ , $\lambda/\mu  = 0.5$. We used a Chebyshev
grid with 25 points.}
\end{figure}

The boundary conditions for the various fields at the UV are
\begin{equation}
	U = 1 + O(z) , \qquad V_1 =  1 + O(z), \qquad V_2 =  1 + O(z) , \qquad \chi = \lambda + O(z), \qquad a = \mu + O(z)
\end{equation}
\noindent while at the IR we impose regularity. We impose this condition by performing a series 
expansion near the horizon assuming that all fields admit regular expansions, and reading off the resulting 
mixed boundary conditions. Following the counting in \cite{Donos:2013eha}, we can argue that
the solutions are specified by 3 dimensionless parameters, $T/\mu$, $k/\mu$, and $\lambda/\mu$. 
Note that $\lambda$ and $k$ play the role of the amplitude and wave number of the lattice, respectively. We show
the profile of $\chi$ in Fig. \ref{fig:chi_QL}.

\section{Linear PDEs}
\label{sec:linearPDE}

There is a significant difficulty leap by adding an extra dimension to the EVPs or
BVPs we have discussed above. To some extent, this is due to the fact that there is 
no simple short cut similar to shooting to solve these problems, and the ``black box" 
type of tools available (e.g. higher dimensional \MT{NDSolve}) are often quite limited. 
However, for the application we are interested in, the domain is rectangular and in that 
case the overall strategy does not differ that much from the 1D discretization method, 
in that we need to define a computational grid and obtain the corresponding derivative operators. 
%

\subsection{Tensor product and bookkeeping}

For rectangular domains, the most efficient way to construct the derivative operators is via the tensor 
product of the 1D derivative operators. This is explained in great detail in \cite{matlab}, and 
the reader is advised to consult it for details. In Mathematica, you can take tensor products with the command
\MT{KroneckerProduct}. 
Moreover, Mathematica has also built-in the discrete derivative operators and it is quite flexible 
with the options, e.g. you can have periodic boundary conditions along one direction and 
Chebyshev pseudospectral on the other. The command you want is \MT{NDSolve`FiniteDifferenceDerivative}. We will see
how to include the appropriate options in the afternoon session.

The case can be made that the main difficulty of this procedure is in the bookkeeping, since 
the so constructed derivative matrices do not act in the ``physical" space of the 2D grid, -- most naturally 
visualized as a matrix -- but in the space of the resulting ``flattened" vectors. This is roughly 
speaking taking the matrix of values on the 2D grid and arranging them as a 1D vector following a particular order.
The opposite procedure, \MT{Partition} in Mathematica, is also very important. This allows us to take the 
flattened vectors and reconstruct our variables in grid form. From this short paragraph it is already 
apparent how confusion can easily ensue.  

Of course both representations, the vectorial and the tensorial one, are equivalent, but there are some operations
which are easier to visualize and code using one or the other. Most notably, imposing the boundary conditions 
is easier to implement in the grid form, since we can simply think of them as being equations of motion 
on their own and then replace the corresponding pieces of the bulk equations of motion lying literally on the edges 
of the operators in grid form. Once the coefficients of the multiplicative operators have been assembled -- including 
the boundary conditions -- we can flatten them with confidence and compose them with the tensor product 
derivative operators. 
We are then left with an matrix problem whose solution is expressed in the flattened form. This can be partitioned
to put it in grid form for visualization or subsequent manipulation. 

\subsection{Example 5: Fiddling around with \texttt{Flatten} and \texttt{Partition}}

At this point things got already quite confusing so an example is in order. We will consider 
the discrete version of the operation
\begin{equation}
\label{PF exercise}
	\sin (x) \sin (y)  \partial_y \partial_x  \cos(x) \cos ( y) =  \sin^2 (x) \sin^2 (y), 
\end{equation}
\noindent on a homogeneous periodic grid $[- \pi , \pi] \times [- \pi , \pi]$. \\

\noindent We begin by defining the grids $\vec x_i = \{ x_i \}$ and $\vec y_i = \{ y_i \}$ in the usual way. 
Given these, we define the derivative matrix in Mathematica specifying a pseudospectral degree and periodic 
boundary conditions, call it $D_{xy}$. 
It is very useful to define the grid-form matrices 
\begin{equation}
	x_t := \vec x \otimes 1, \qquad y_t := 1 \otimes \vec y
\end{equation}
These correspond to functions defined on the grid which take the value $x$ for all $y$ and viceversa. The 
subindex $t$ stands for tensor. 

Using these tensors it is really easy to define with Mathematica the matrices that 
represent the functions which act as multiplicative factors of the derivative operators: 
we simply evaluate a function $u(x, y)$ using {\it simultaneously} the replacement $\{ x \to x_t , y \to y_t \} $. 
In our example, this generates the grid-form matrices $SS$ and $CC$ representing the products of sines and cosines. 
In order point-wise multiply the derivative operator, we flatten them obtaining $F[SS]$, $F[CC]$, 
and take the Hadamard product $O:= F[SS] * D_{xy}$. 
This operator $O$ acts on objects of the flatten type, so we can use standard matrix multiplication to compose it with 
$F[CC]$, resulting in the flattened vector $O. F[CC]$. This can be brought into the grid form by taking its partition, $P[ O. F[CC] ]$. 

\begin{figure}[h]
\centering
\includegraphics[scale=0.5]{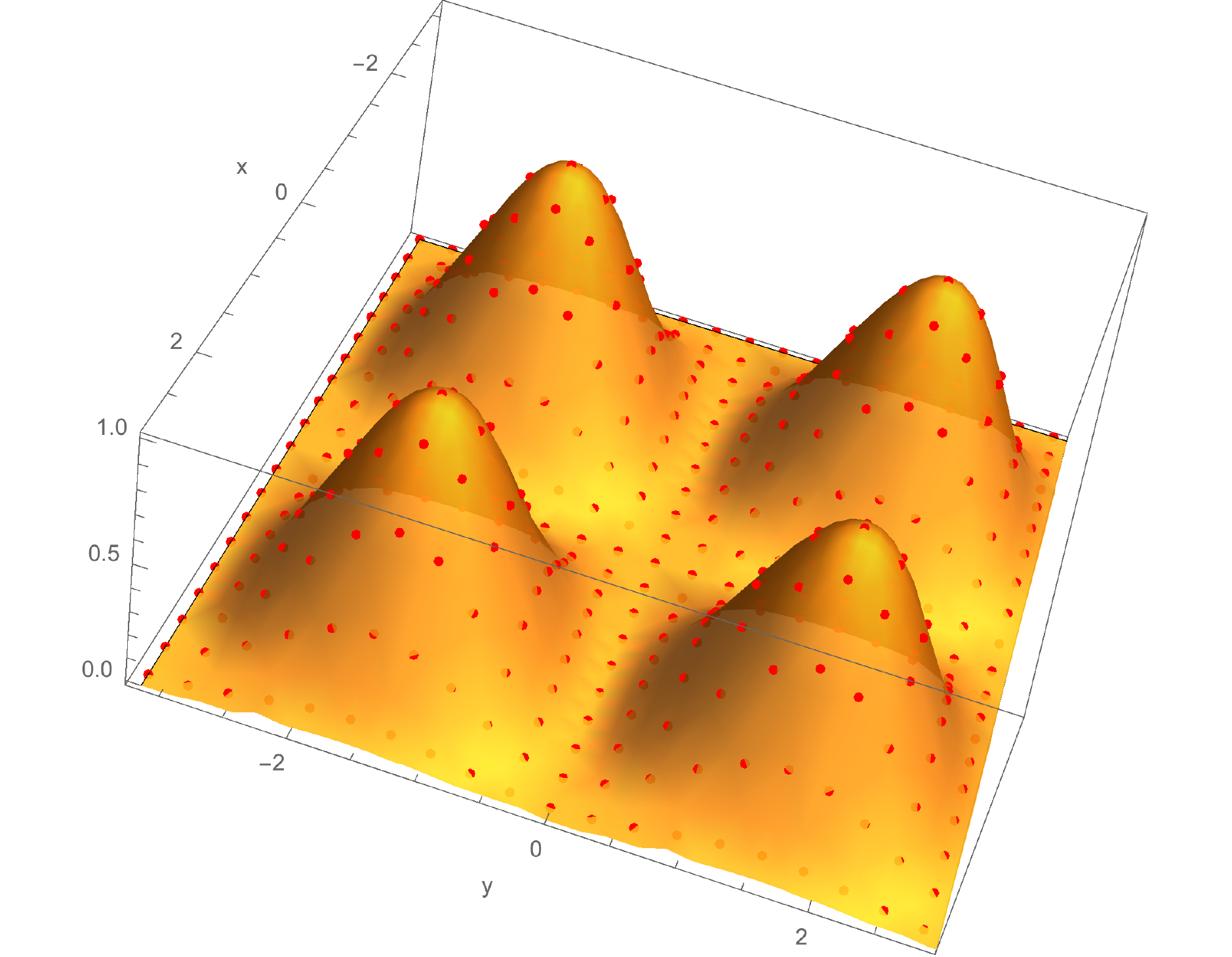}
\caption{\label{fig:discrete derivative} The points show the discrete version of the l.h.s of \eqref{PF exercise} computed 
on a homogeneous grid with periodic pseudospectral differentiation matrices, while the surface is the r.h.s. of \eqref{PF exercise}.}
\end{figure}

In Fig. \ref{fig:discrete derivative} we compare the discretized version of the left hand side of \eqref{PF exercise} with the analytic expression 
in the right hand side, finding excellent agreement. We strongly advise our readers to become familiar with all these manipulations. 

\subsection{Example 6: Poisson equation}

We consider the problem (eq. (7.4) in \cite{matlab})
\begin{equation}
\label{Poisson BVP}
	\partial_x^2 u + \partial_y^2  u = 10 \sin [ 8 x (y-1) ], \qquad - 1 < x , y < 1 , \qquad u = 0 \quad  {\rm on \, the \, boundary}
\end{equation}
In order to treat the bulk equations and the boundary conditions on the same footing, it is convenient to consider the general problem 
\begin{equation}\label{general linear PDE}
	( \phi_{xx}(x,y) \partial_{xx} + \phi_{xy}(x,y) \partial_{xy} + \phi_{yy}(x,y) \partial_{yy} + \phi_{x}(x,y) \partial_{x} + 
	\phi_{y}(x,y) \partial_{y} + \phi_0 (x,y)  ) u = j(x,y)
\end{equation}
We have 6 different derivative operators, and we need to identify their coefficients. 
This can be done in Mathematica simply by taking partial derivatives of the equation of motion  wrt $ \partial_{xx}  u$, 
etc. It is easy to discretize these using the replacements $\{ x \to x_t , y \to y_t \} $ as discussed in the previous section. 
In order to treat the boundary conditions, we can extend them to all the domain and and extract the coefficients in the same 
way as we did in the bulk\footnote{Although conceptually clear, this strategy is not very efficient in terms of memory consumption since it requires us to store
lots of unnecessary data corresponding to the boundary conditions evaluated in places where they are meaningless.}. 
In this way, it is easy to replace the boundary conditions in the appropriate places when all the functions $\phi_{ij}(x, y)$ are still 
in their grid form. 

After obtaining the $\phi_{ij}(x, y)$ with the {\it boundary conditions implemented}, we flatten them and Hadamard 
multiply them with the respective matrix derivative operators. We can then add them all together obtaining 
a matrix that needs to be inverted to yield the solution of our problem by standard matrix multiplication with 
the discrete and flattened version of $j(x,y)$. 

\begin{figure}[h]
\centering
\includegraphics[scale=0.5]{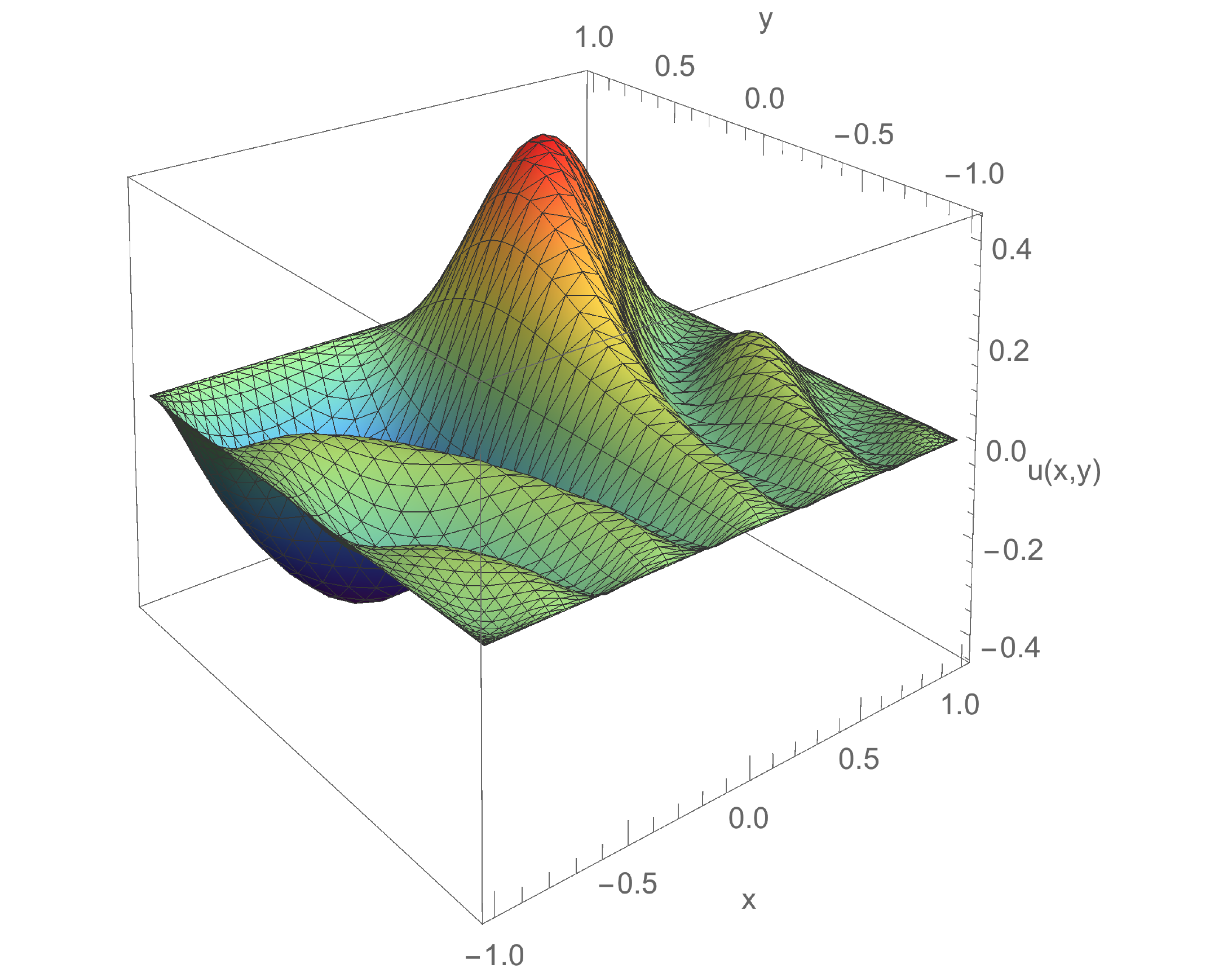}
\caption{\label{fig:soln_Poisson} Solution to the Poisson eq \eqref{Poisson BVP} with a $40 \times 40$ grid of Chebyshev and pseudospectral differentiation.}
\end{figure}

The problem \eqref{general linear PDE} clearly fits in this general class of problems, so we can solve them with the method
described in this section. We plot the solution in Fig \ref{fig:soln_Poisson}.

\section{Non-Linear PDEs}
\label{sec:NLPDE}

We now turn to the hardest problem of our series, namely BVPs involving non-linear PDEs. As we shall 
see, this setup is relevant in the construction of black holes dual to inhomogeneous 
holographic lattices. 

Most of the ingredients to tackle this challenge are in place: we know how to solve 
BVPs involving linear PDEs with arbitrary sources and boundary conditions. As in the 1D case, 
the solutions of interest can be obtained by applying a NR procedure to the non-linear system of interest. 

There is however one crucial point which is present in the gravitational setup and that we need to address: because of 
diffeo invariance, the operators that we obtain when linearizing Einstein's equations around a given configuration 
contain zero modes, so that the boundary value problem is not elliptic and as such it does not converge. We will discuss 
how to circumvent this issue in the next section. 

\subsection{DeTurk trick}

Here is a clever trick originally due to DeTurk: in order to restore ellipticity, we simply add a term, denoted as the ``DeTurk term" ,
to Einstein's equations which makes them explicitly elliptic at {\it every step}. We need to make sure however that we
are actually solving Einstein's equations! This is the most interesting part of the ``trick": under certain assumptions, 
we can show that the DeTurk term we added must converge to zero once a solution is reached. Moreover, for configurations in which 
the NR algorithm converges and the DeTurk term survives -- called Ricci solitons--, can be shown to be isolated points in 
the space of solutions, so they are easily distinguishable from true solutions to Einstein's equations. This method 
was originally used to obtain stationary black hole solutions in \cite{Headrick:2009pv, Adam:2011dn, Wiseman:2011by}, 
see \cite{Dias:2015nua} for a review. 

Let us discuss in detail how this works. The first step is to identify a {\it reference metric} $\bar g_{\mu \nu}$, which 
is a geometry satisfying the same boundary conditions as the solutions we would like to find. We then construct the Christoffel 
symbol of the reference metric, $\bar \Gamma^\mu_{\nu \rho}$, and define the vector
\begin{equation}
	\xi_\nu = (\Gamma^\mu_{\nu \mu} - \bar \Gamma^\mu_{\nu \mu})
\end{equation}
\noindent where $\Gamma^\mu_{\nu \rho}$ is the Christoffel symbol of the metric we are solving for. The Einstein equation is then 
modified by replacing
\begin{equation}
	R_{\mu \nu} \to R_{\mu \nu}^H :=  R_{\mu \nu} + \nabla_{(\mu} \xi_{\nu)} 
\end{equation}
The quantity $R_{\mu \nu}^H$ is known as the harmonic Ricci tensor, and it has the property that its 
linearization with respect to an arbitrary background metric is manifestly elliptic. 

In order to setup the NR algorith properly, we need to allow for the most general metric compatible 
with the symmetries of the problem {\it without} fixing the gauge by hand. One can think of the convergence of the 
algorithm to a solution with $\xi^\mu \xi_\mu = 0$ (which is a complicated function of the metric functions) 
fixes the gauge dynamically as we move around field space. 
In practice, what we do is to setup the NR algorithm for the modified Einstein equations and check that 
the norm of $\xi$ vanishes on the solutions. As mentioned above, this is guaranteed in some simple cases, 
but in general this property could fail so we need to verify it explicitly. 

Having introduced this new trick, we are now in a good position to discuss the construction 
of inhomogeneous holographic lattices. 

\subsection{Example 7: Holographic Ionic lattice}

Our final example consists of a black hole solution with modulated chemical potential. These were 
first found in \cite{Horowitz:2012gs}, based on the previous work \cite{Horowitz:2012ky}. We closely follow 
these references in our presentation. 
Let us consider the Einstein-Maxwell system with negative cosmological constant in four dimensions, 
\begin{equation}
	I = \int d^4 x \sqrt{- g} \left( R + 6 - \frac{1}{4} F^2  \right)
\end{equation}
The holographic dictionary maps the leading term of $A_t$ to the chemical potential $\mu$. 
The configurations we are after break translations explicitly by introducing spatial 
dependence in $\mu$. For simplicity, we will restrict ourselves to solutions in which 
$\mu$ depends only on one boundary coordinate $x$. Therefore, the functions we are solving 
for contain dependence on the holographic coordinate $z$ and the boundary coordinate $x$. 
As described above, using the DeTurk trick we can render the so obtained eoms elliptic so we can indeed 
apply the NR method to find the desired solutions. 

\begin{figure}[h]
\centering
\includegraphics[scale=0.5]{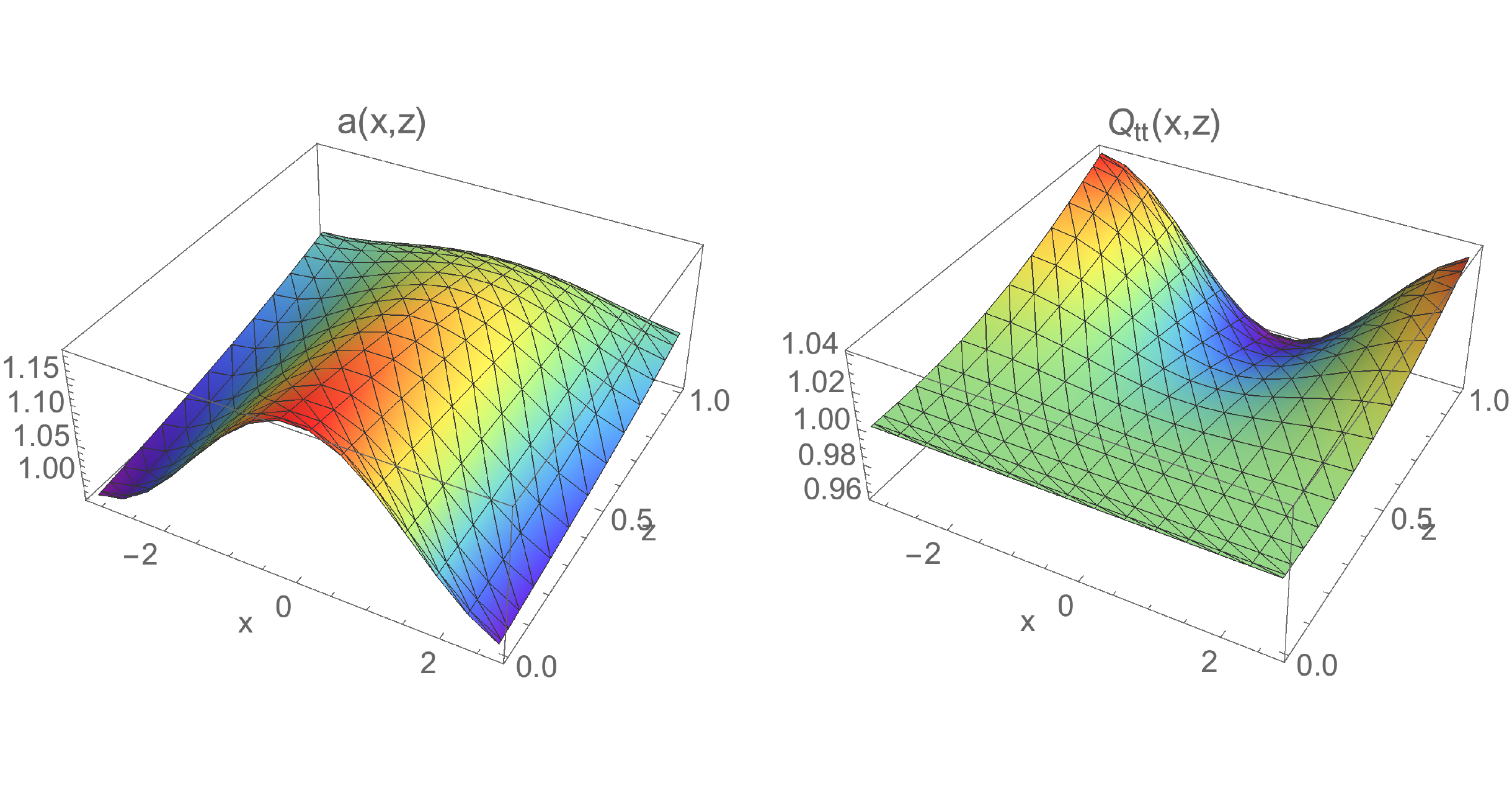}
\caption{\label{fig:ionic} Profiles of $a$ and $Q_{tt}$ for the ionic lattice with $T/\mu =0.2$, $k/ \mu = 1$, $A_0 = 0.1$}
\end{figure}

The following ansatz suffices to accommodate the solutions of interest
\begin{align}
 	ds^2 &= \frac{1}{z^2} \left[ - f(z) Q_{tt} dt^2 + \frac{Q_{zz}}{f(z)} dz^2 + Q_{xx}( dx + Q_{xz} dz )^2 + Q_{yy} dy^2  \right] \\
 	A &= (1- z) a dt
\end{align} 
\noindent where 
\begin{equation}
	f = (1-z)(1 + z + z^2 - \bar \mu^2 z^3/ 4)
\end{equation}
We obtain the RN solution by letting $Q_{tt} = Q_{zz} = Q_{xx} = Q_{yy} = 1$, $Q_{xz} = 0$, 
$a = \bar \mu (1-z) $. The temperature of this solution is
\begin{equation}
\label{Tovermu}
	T = \frac{12 - \mu^2}{16 \pi}
\end{equation}
We source a monochromatic lattice by introducing a modulated chemical potential by considering the boundary conditions 
\begin{equation}
	a = \mu (x) \qquad {\rm at} \quad z = 0
\end{equation}
\noindent with
\begin{equation}
	\mu (x) = \mu_0 (1 + A_0 \cos ( k x )  )
\end{equation}
The boundary conditions for the metric are AdS asymptotics 
\begin{equation}
	Q_{tt} = Q_{zz} = Q_{xx} = Q_{yy} = 1 , \qquad Q_{xz} = 0 , \qquad {\rm at} \quad z = 0
\end{equation}
We require regularity at the horizon, which results into the condition $Q_{tt}(x, 1) = Q_{zz} (x, 1)$ in addition 
to other five boundary conditions of mixed type for the remaining functions. 
Regular solutions have a constant temperature across the horizon (as required by the zeroth law of black hole thermodynamics) which is 
given by \eqref{Tovermu}. 
Without loss of generality, we can set $\mu_0 = \bar \mu$ and dial the dimensionless ratio $T/\mu$ by varying $\bar \mu$. 
We expect to obtain a family of solutions which depend on 3 dimensionless parameters, $A_0$, $k\mu$ and $T/\mu$. Implementing the NR algorithm we are able to find the solution depicted in Fig \ref{fig:ionic}. 

\begin{figure}[h]
\centering
\includegraphics[scale=0.5]{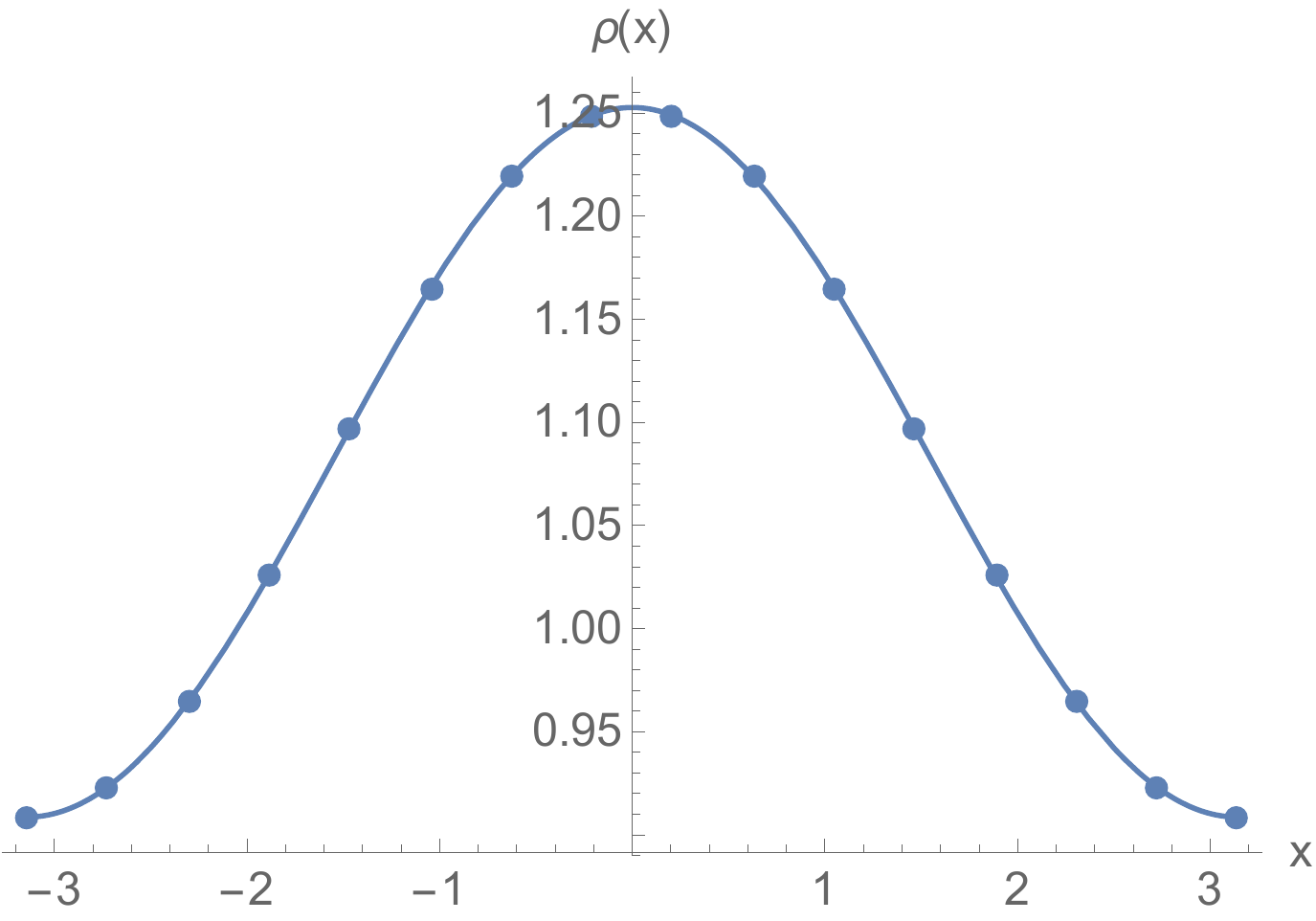}
\caption{\label{fig:rho_ionic} Charge density of the ionic lattice with $T/\mu =0.2$, $k/ \mu = 1$, $A_0 = 0.1$}
\end{figure}

In order to extract the one point functions, 
we need to read off the subleading terms in the boundary expansion. For example, the charge density $\rho(x)$ is given by 
\begin{equation}
	A_t = \mu (x) - z \rho(x) + O(z^2)
\end{equation}
We plot the charge density in Fig \ref{fig:rho_ionic}.

A very important property of these lattices is that the optical conductivity can be approximated by a Drude peak 
at small frequencies, which means in particular that their DC conductivity is finite. Extracting the optical conductivity
requires perturbing the solutions above, which yields to a large set of fluctuation equations containing gauge redundancies. This 
has been done in the literature \cite{Horowitz:2012gs, Horowitz:2012ky, Donos:2014yya, Rangamani:2015hka}, 
but it is out of the scope of this lectures. We invite the enthusiastic students to 
take up this challenge.

\section{Outlook }

In these series of lectures we have discussed some of the essential technical tools for researchers aiming 
to tackle problems in the area of what could be called ``applied holography". Our main goal was to construct 
black hole solutions breaking translational invariance along the boundary directions inhomogeneously. 
Along the way, we developed important numerical techniques, which are not only useful to holography, but also 
to physics and applied mathematics more broadly. 

Due to time constraints, I had to leave out many interesting topics and applications which I would like to 
suggest for future study. Here is an incomplete list:

\begin{itemize}

\item Construction of other homogeneous black hole solutions which break translations explicitly, such as the helical lattices. 
As in the case of the Q-lattice, these can be found solving a system of non-linear ODEs. 

\item Construction of solutions which break symmetries spontaneously, such as the holographic 
superconductors of HHH, or the charge density waves of \cite{Nakamura:2009tf, Donos:2012wi}. This is facilitated by first finding the perturbative 
condensate solution and using this as a seed to feed the non-linear problem. 

\item Study linearized perturbations around black hole backgrounds which break translations explicitly. A particularly simple
starting point is the Einstein-axion model of \cite{Andrade:2013gsa}, in which case the background is analytic and the perturbation equations can be decoupled. The optical conductivity presents the Drude behaviour characteristic of configurations which relax momentum.

\item A more challenging problem is to perturb numerical black holes. When these break translations, usually one obtains a large 
number of perturbation equations and extracting the gauge invariant information is a non-trivial task. 

\item There is an interesting formalism to study time evolution in asymptotically AdS spaces, which roughly 
speaking consists of solving nested ODEs at each step of the time integration \cite{Chesler:2008hg, Chesler:2013lia}. This is enormously simplifies the problem of dynamical  evolution, since we can use most of the techniques applicable to static problems to investigate time-evolving geometries. 

\item We have not given the space it deserves to the study of accuracy of our numerics. We have touched upon the issue 
of convergence with increasing number of grid points, but this cannot continue indefinitely due to the fact that large matrices 
also introduce large rounding errors. There is an ``optimal range" of grid size, and this can be estimated by studying how the 
extracted physical quantities depend on it.

\item We have only considered monochromatic lattices with a single wave number in the chemical potential. There are interesting effects
which take place when we consider more wave number components, the extreme case being a disordered lattice. 

\end{itemize}

\section*{Acknowledgements}

I wrote these notes based on the numerical techniques I have learned in the past six years. Many 
people have shared with me their knowledge and wisdom in this subtle subject, and I am 
immensely grateful to all of them. 
In particular, it is a pleasure to thank Alexander Krikun, Jorge Santos and Benjamin Withers, who have patiently and 
generously guided me through many aspects of this subject. 
My work is supported by the ERC Advanced Grant GravBHs-692951.
My participation at this school was funded by the Newton-Picarte Grant 20140053, the support of which I 
gratefully acknowledge. I would like to thank the organizers, specially Julio Oliva, for their hospitality, 
and the students for their sharp questions and enthusiasm.

\bibliographystyle{JHEP-2}
\bibliography{Notes_NumHol}

\end{document}